\documentclass[printer]{aa}
\usepackage[varg]{txfonts}
\usepackage{graphicx}
\usepackage{booktabs}
\usepackage{threeparttable}
\usepackage{natbib}
\newcommand{\kms}{km~s$^{-1}$}
\newcommand{\FeXXI}{\ion{Fe}{xxi}}
\newcommand{\CI}{\ion{C}{i}}
\newcommand{\CII}{\ion{C}{ii}}
\newcommand{\SI}{\ion{S}{i}}
\newcommand{\MgII}{\ion{Mg}{ii}}
\newcommand{\SiIV}{\ion{Si}{iv}}
\newcommand{\SiII}{\ion{Si}{ii}}
\newcommand{\FeII}{\ion{Fe}{ii}}
\newcommand{\OI}{\ion{O}{i}}
\newcommand{\OIV}{\ion{O}{iv}}
\newcommand{\FeVIII}{\ion{Fe}{viii}}

\newcommand{\FeXXIII}{\ion{Fe}{xxiii}}
\newcommand{\FeXII}{\ion{Fe}{xii}}

\begin{document}

\title{Observations of solar flares with IRIS and SDO}
\author{D. Li \inst{1,2,3}, D.E. Innes \inst{2}, and Z. J. Ning \inst{1}}
\institute{Key Laboratory of Dark Matter and Space Science, Purple Mountain Observatory, CAS, 210008 Nanjing, China (\email{lidong@pmo.ac.cn}) \\
               \and Max-Planck Institute for Solar System Research, 37077 G\"{o}ttingen, Germany\\
               \and University of Chinese Academy of Sciences, 100049 Beijing, China}
\date{Received; accepted}

\abstract{Flare kernels brighten simultaneously in all Solar
Dynamics Observatory (SDO) Atmospheric Imaging Assembly (AIA)
channels making it difficult to determine their temperature
structure. The Interface Region Imaging Spectrograph (IRIS) is able
to spectrally resolve \FeXXI\ emission from cold chromospheric
brightenings, so can be used to infer the amount of \FeXXI\ emission
in the 131~\AA\ AIA channel.} {We use observations of two small
solar flares seen by IRIS and SDO to compare the emission measures
(EMs) deduced from the IRIS \FeXXI\ line and the AIA 131~{\AA}
channel to determine the fraction of \FeXXI\ emission in flare
kernels in the 131~\AA\ channel of AIA.} {Cotemporal and cospatial
pseudo-raster AIA images are compared with the IRIS results. We use
multi-Gaussian line fitting to separate the blending chromospheric
emission so as to derive \FeXXI\ intensities and Doppler shifts in
IRIS spectra.} {We define loop and kernel regions based on the
brightness of the 131 and 1600~\AA\ intensities. In the loop regions
the \FeXXI\ EMs are typically 80\% of the 131~\AA\ ones, and range
from 67\% to 92\%. Much of the scatter is due to small misalignments
but the largest site with low \FeXXI\ contributions was probably
affected by a recent injection of cool plasma into the loop. In
flare kernels the contribution of \FeXXI\ increases from less than
10\% at the low intensity 131~\AA\ sites to 40-80\% in the brighter
kernels. Here the \FeXXI\ is superimposed on bright chromospheric
emission and the \FeXXI\ line shows blue shifts, sometimes extending
up to the edge of the spectral window, 200~\kms.} {The AIA 131~\AA\
emission in flare loops is due to \FeXXI\ emission with a 10-20\%
contribution from continuum, \FeXXIII, and cooler background plasma
emission. In bright flare kernels up to 52\% of the 131~\AA\ is from
cooler plasma. The wide range seen in the kernels is caused by
significant structure in the kernels which is seen as sharp
gradients in \FeXXI\ EM at sites of molecular and transition region
emission.}


\authorrunning{Li et al.}
\titlerunning{Observations of solar flares with IRIS and SDO}

\maketitle

\section{Introduction}
The flare impulsive phase is characterized by a sudden increase in
chromospheric and hard X-ray emission \citep{Fletcher01,Fletcher11}.
The emission is concentrated in small bright kernels and along
ribbons, coinciding with magnetic field concentrations, where rapid
chromospheric heating  drives hot plasma upward into the corona
\citep{Mason86,Teriaca06,Benz08,Ning11}. The kernels are thought to
be the chromospheric signature of magnetic reconnection in the
corona \citep{Qiu02,Fletcher04,Gan08}. Spectroscopic observations
have revealed  high-temperature, high-velocity blue and cooler
red-shift emission \citep{Antonucci82,Milligan06,Teriaca06,Young13}
compatible with models of chromospheric evaporation.

Observations of the kernels show simultaneous brightening in all AIA
extreme ultraviolet (EUV) channels
\citep{Brosius12,Fletcher13,Young13}. The analysis by
\citet{Brosius12} that compared AIA images with CDS spectra, showed
that in a small GOES B4.8 flare a significant fraction of the hot
channel, 94~\AA\ and 131~\AA, emission could be attributed to
brightening of transition region and lower coronal lines. On the
other hand \citet{Fletcher13} attributed all the 131~\AA\
brightening in an M1.0 flare to plasma with temperatures greater
than 10~MK. In an M1.1 flare, \citet{Young13}'s analysis of EIS
spectra  found almost equal emission measures (EMs) across the
observed temperature range from 0.1 to 10~MK but did not discuss the
contributions to the AIA channel images.

The main contribution to the 131~\AA\ channel in flares is the
\FeXXI\ 128.75~\AA\ line \citep{ODwyer10,Milligan13}. The same ion,
\FeXXI\ produces the strong flare line at 1354.08~\AA\ in IRIS
spectra. Therefore by comparing simultaneous IRIS and AIA
observations it may be possible to determine the contribution of
\FeXXI\ to the 131 channel and hence resolve the the question of the
cool plasma emission contribution to the 131 channel. Using the same
ion, has the advantage that the line ratios are independent of
ionization and only weakly dependent on temperature.

The forbidden line of \FeXXI\ at 1354.08 \AA\ has been used in
several spectroscopic studies to investigate hot plasma flow during
flares
\citep{Doschek75,Cheng79,Mason86,Feldman00,Innes01,Kliem02,Wang03}.
More recently \citet{Young15} described high spatial and spectral
resolution IRIS observations of \FeXXI\ from hot flare kernels and
loops, with temperatures about 10~MK from an X1 class flare. Their
results support the chromospheric evaporation model. The
\citet{Mason86} observations indicated that the \FeXXI\ line is
often blended with \CI. The higher resolution spectra analysed by
\citet{Young15} show that the \FeXXI\ kernel emission may, in
addition, be blended with other chromospheric and possibly molecular
lines. Molecular hydrogen was identified at the footpoints of X-ray
loops with SUMER spectra \citep{Innes08}. To obtain estimates of the
\FeXXI\ EM, the strength of these chromospheric lines needs to be
taken into account. For the analysis in this paper, we have
investigated a number of flare kernel spectra and developed an
algorithm for simultaneously fitting the \FeXXI, blending lines, and
continuum.

The EM obtained from the IRIS 1354~\AA\ line was compared with that
derived from simultaneous SDO/AIA 131~\AA\ images to determine the
contribution of cooler line emission to the 131~\AA\ channel at the
site of flare kernels. We find that the \FeXXI\ observed by IRIS can
account for about 80\% of the SDO/AIA 131~\AA\ emission in flare
loops and 40-80\% of the emission in flare kernels. Assuming that an
additional 20\% of the \FeXXI\ is continuum \citep{Milligan13}, we
find that up to 52\% of the 131~\AA\ channel emission is from cooler
plasma in the flare kernels.

\section{Observations}
\begin{figure} %%%%%%%%%%%%%%%%%%
\centering
\includegraphics[width=\linewidth,clip=]{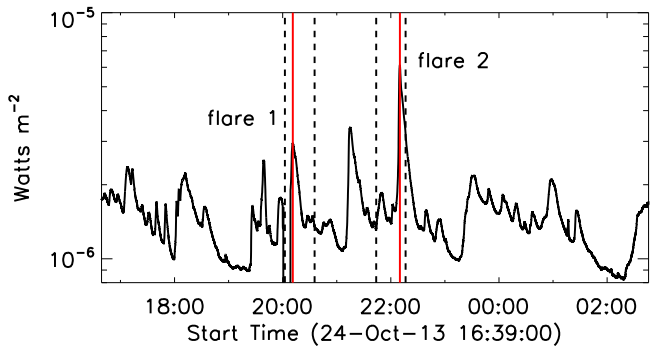}
\caption{GOES 1.0-8.0~\AA\ flux during the observing period. The two
flares analysed are labelled. Red lines are drawn at the times of
the peak flux, while the dashed lines show the start and end times
of the of IRIS raster~6 (between 20:02 UT and 20:35 UT) and raster~9
(from 21:43 UT to 22:16 UT).} \label{fig_goes}
\end{figure}

\begin{figure} %%%%%%%%%%%%%%%%%%
\centering
\includegraphics[width=\linewidth,clip=]{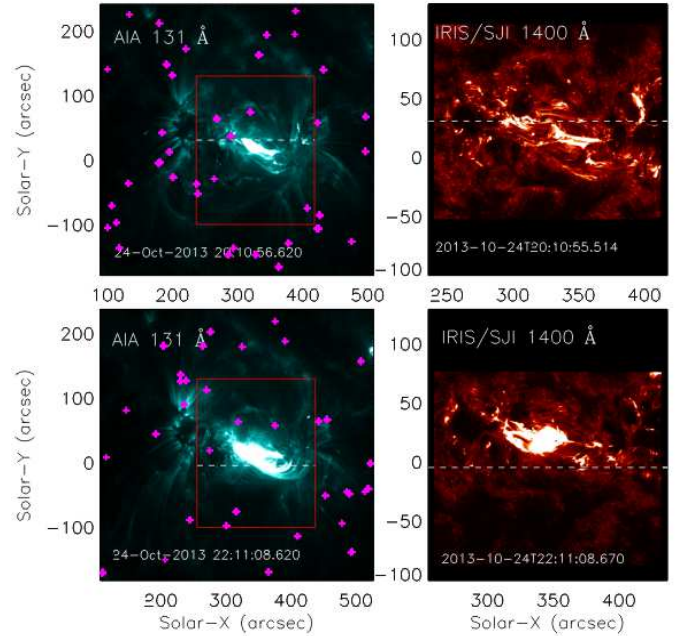}
\caption{Left: AIA 131~\AA\ images of the active region at the times
of the two flares. The red boxes are the positions of the IRIS SJ
images. Right: IRIS SJ 1400~\AA\ images of the two flares analysed.
The position of the spectrometer slit at the start of the flare is
seen as a white dashed horizontal line. The purple pluses mark the
re-spike pixels in the AIA 131~\AA\ images with the code `aia\_
respike'.} \label{fig_over}
\end{figure}

The observed active region, AR 11875, produced four C-class flares
during the period 16:39~UT on 24 October 2013 and 02:46~UT on 25
October 2013. We obtained high quality cotemporal and cospatial IRIS
and AIA observations for flares at 20:10 and 22:05~UT
(Fig.~\ref{fig_goes}). During the early part of the sequence SDO was
off-pointing and during the 21:09~UT flare, the IRIS spectra were
badly affected by a large number of particle hits.
Fig.~\ref{fig_over} shows the AIA 131~\AA\ and corresponding
slit-jaw (SJ) 1400~\AA\ images of the two flares analysed. The GOES
fluxes of the two flares are shown in Fig.~\ref{fig_goes}. Flare~1
peaked at about 20:10 UT and ended at about 20:22 UT, while the
flare~2 peaked at about 22:10 UT and stopped at about 22:15 UT.

\subsection{AIA observations}
In this analysis, AIA images \citep{Lemen12} from the 131 and
1600~{\AA} channels are used to compare with the IRIS data. The AIA
level 1.0 data were downloaded and processed to level 1.5 using the
standard solarsoft (SSW) routines. Then using the code
`drot\_map.pro' in SSW, the active region with a FOV of 420\arcsec
$\times$ 420\arcsec are selected from the AIA full disk images
(see., Fig~\ref{fig_over}). The images from different AIA filters
have alignment uncertainties of about 1$-$2 pixels
\citep[e.g.,][]{Young13}. Therefore the AIA 131~\AA\ images are
slightly shifted (i.e., 0.5$-$2 pixels) with respect to the
1600~\AA\ images because we obtained the best alignment of the
1600~\AA\ with the IRIS 1400~\AA\ and the 131~\AA\ with the IRIS
\FeXXI.

The downloaded AIA data are already de-spiked which often removes
flare kernel emission and these can be put back with the
aia\_respike routine \citep[see.,][]{Young13}. In our study, we have
compared the `re-spike' data and the level 1 data at 131~\AA\, as
shown with the purple pluses in Fig.~\ref{fig_over}. This shows that
the re-spiked pixels are far away from the IRIS slit position, and
they are not within the regions of our study. This was the case for
all other 131~\AA\ images. Also during big flares the AIA 131~\AA\
channel is often saturated. The two flares in this paper are quite
small and only very few pixels in the 131~\AA\ channel were
saturated. Fortunately, those saturated pixels are far away from the
IRIS raster slit. In other words, neither the de-spiking nor the
saturated pixels are a problem in our study.

\begin{figure} %%%%%%%%%%%%%%%%%%%%%%%%%
\centering
\includegraphics[width=\linewidth,clip=]{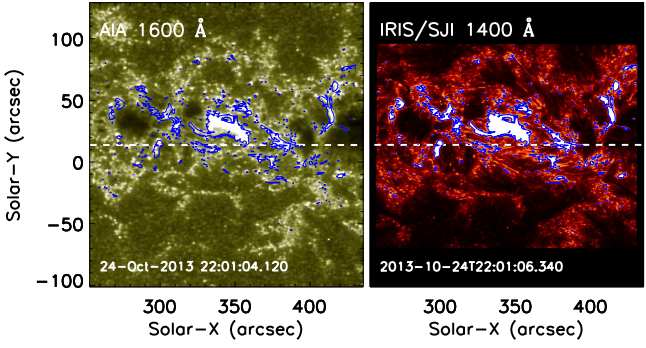}
\caption{Coalignment of the SDO and IRIS images. Left: the AIA image
at 1600~\AA. Right: the SJ image at 1400 {\AA}. The white dashed
line is the IRIS slit position, and the blue contours are the SJ
intensity at the level of 800~DN.} \label{fig_coalign}
\end{figure}

\subsection{IRIS observations}
IRIS is a NASA Small Explorer Mission launched in June 2013, and its
main science is an investigation of the dynamics of the Sun's
chromosphere and transition region \citep{McIntosh13}. As described
in detail by \citet{Depontieu14}, IRIS is a high resolution
spectrograph and slit-jaw (SJ) camera which obtains spatial
resolution of 0.33-0.4 arcsec, and spectral resolution of
$\sim$26~m\AA\ (or $\sim$52~m\AA\ in second order). There are three
IRIS wavelength bands: (i) 1332-1358~\AA\ which includes the strong
\CII\ doublet, \FeXXI, and a number of strong \CI\ lines, (ii)
1389-1406~\AA\ which includes the \SiIV\ doublet, and density
sensitive \OIV\ lines, (iii) 2782-2834~\AA\ which includes \MgII\ h
and k lines. Here we analyse lines in the 1332-1358~\AA\ range. IRIS
also obtains SJ images centered at 1330, 1400, 2796, and 2832~\AA.

The observations described here were designed with the aim of
capturing the structure and spectra of flares by making rapid
rasters across a flaring active region for as long a period as
telemetry constraints allowed. The IRIS rasters were obtained by
taking 64 1.01\arcsec\ steps across a region 174\arcsec x 63\arcsec
with the roll angle $90\,^{\circ}$ (i.e. the slit was oriented E-W).
Simultaneous SJ images at 1400~\AA\ with a field-of-view
174\arcsec x166\arcsec\ and cadence 32~s were also obtained and used
mainly for coalignment with 1600~\AA\ AIA images.
Because continuum emission from the temperature minimum is dominant
in many of the bright features seen in both sets of images,
coalignment is possible to within the AIA
pixel size, 0.6\arcsec (Fig.~\ref{fig_coalign}).

The step cadence was 31.6~s and the exposure time 30~s, thus the
raster cadence was 33~m 42~s. 18 rasters were obtained in the
roughly 10 hour period. The pixel size along the slit was at the
highest resolution 0.167\arcsec\ but to save telemetry four times
spectral binning and a restricted number of spectral windows were
obtained. In this case we used the `flare' list of lines which
consisted of the `\CII, 1343, \FeXII, \OI, \SiIV' FUV windows. Here
we only discuss spectra from windows in the short wavelength FUV
band. The spectral resolution was 4*12.72~m\AA/pixel, equivalent to
10.88~\kms/pixel.

\begin{figure}    %%%%%%%%%%%%%%%%%%
\centering
\includegraphics[width=\linewidth,clip=]{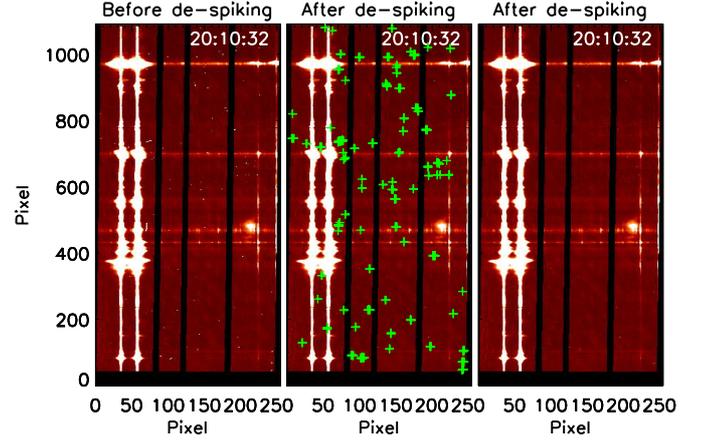}
\caption{IRIS raster images before (left) and after despiking
(right). The middle panel is an image after despiking with the bad
pixels marked with green plus (`+') signs.} \label{despike}
\end{figure}

IRIS level 2 data were downloaded. These data  have been calibrated
and corrected for image distortions. Several of the images were
badly affected by particle hits and hot or dead pixels. These were
removed using a despiking routine that detected persistent hot/dead
pixels, and sudden changes in intensity since these can seriously
distort the line fitting results. Care is taken to ensure that only
isolated bright pixels or small pixel groups are removed, and to
leave intensity changes due to sudden line broadening.
Fig.~\ref{despike} shows the effect of despiking. In addition, small
spectral shifts caused by thermal drifts and the spacecraft orbital
velocity were corrected with the routine
iris\_orbitvar\_corr\_12.pro \citep{McIntosh13,Tian14,Cheng15} in
the SSW package.

\subsection{Flare ribbon spectra}
\begin{figure}    %%%%%%%%%%%%%%%%%%
\centering
\includegraphics[width=\linewidth,clip=]{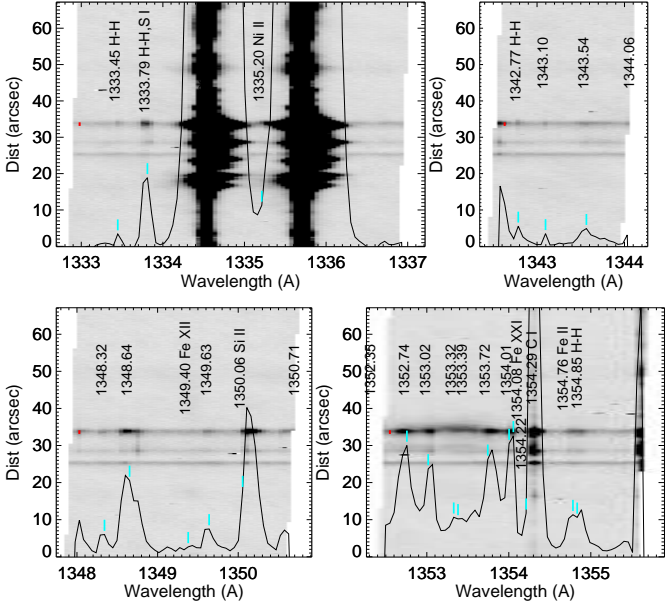}
\caption{Flare footpoint spectra taken on 2013 October 24. The
overplotted line spectra are the average across the small bright
region marked by a red line on the lefthand side of each image. The
main lines are labelled and indicated by light blue vertical ticks
just above the line spectra.} \label{linesoct24}
\end{figure}

\begin{figure}
\centering
\includegraphics[width=\linewidth,clip=]{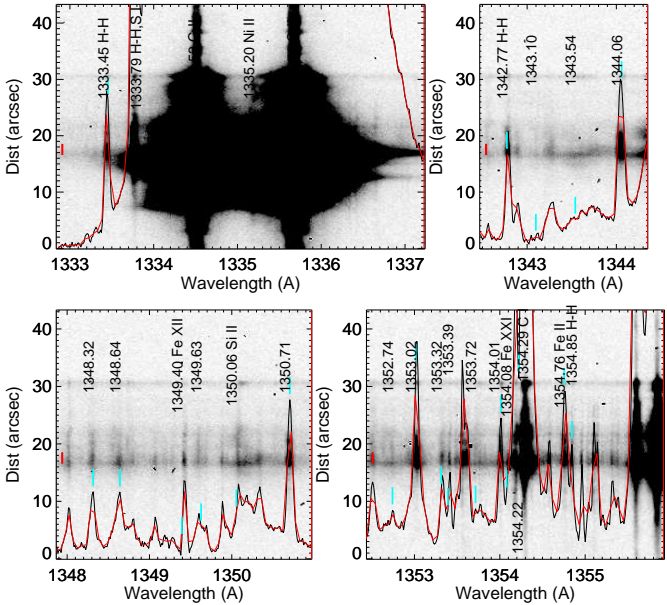}
\caption{Flare footpoint spectra taken on 2013 October 12 at full
resolution. The overplotted line spectra are the average across the
small bright region marked by a red line on the lefthand side of
each image, and, in red, the same spectra binned by a factor 4 which
is the resolution used on 2013 October 24.} \label{linesoct12}
\end{figure}

The ribbons are characterized by many narrow, bright emission lines.
The \FeXXI\ emission blends with both known and unknown lines from
neutral and singly ionized species, as well as molecular
fluorescence lines. To determine the \FeXXI\ intensity we need to
extract the blending chromospheric emission. We have identified the
main blending lines by looking at kernel spectra from other IRIS
datasets at higher spectral resolution, as well as many spectra in
this dataset to find out which other observed lines behave similarly
to the chromospheric lines in the \FeXXI\ window. An example of a
flare footpoint spectrum taken with this sequence is shown in
Fig.~\ref{linesoct24}. For comparison, the chromospheric lines seen
during a C1.5 flare from AR11861 (-133\arcsec, -285\arcsec) at 00:46
UT on 2013 October 12, taken with the full spectra resolution, are
shown in Fig.~\ref{linesoct12}.

\begin{figure}    %%%%%%%%%%%%%%%%%%
\centering
\includegraphics[width=\linewidth,clip=]{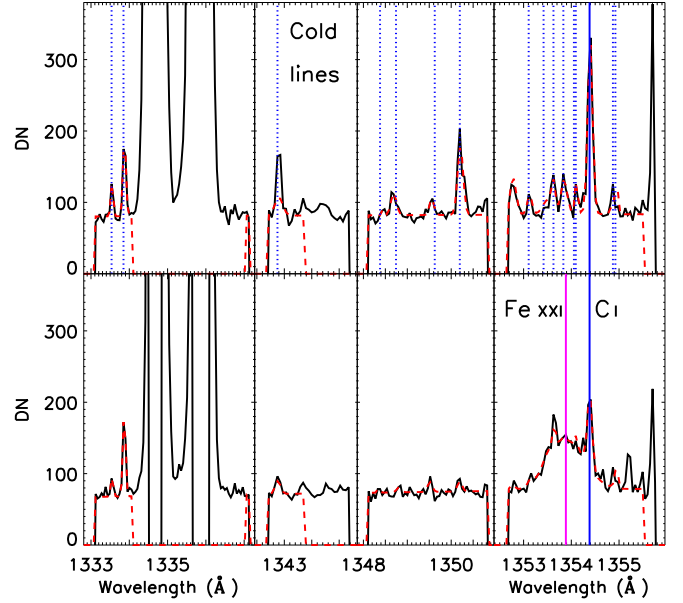}
\caption{The spectra (black) and the multi-Gaussian fits (red). The
blue dashed lines mark the positions of cold lines, the blue solid
line marks the \CI\ line, while the purple line marks the \FeXXI\
line.} \label{goodfits}
\end{figure}

The main chromospheric lines in the `\OI' window, apart from the
\CI\ 1354.29~\AA, are the \FeII\ lines at 1353.02, 1354.01, and
1354.76~\AA, the \SiII\ lines at 1352.74 and 1353.72~\AA, and the
unidentified lines at 1353.32 and 1353.39~\AA. Actually, other
papers \citep[e.g.,][]{Graham15,Li15,Tian15} have attempted
multi-Gaussian fits to the \FeXXI\ line from IRIS spectral
observations. In this paper, to obtain the \FeXXI\ intensities, we
have fixed or constrained the positions and widths of these lines
and set their intensities to have  a specified ratio to
well-resolved lines from similar species. In total we fit 17
Gaussian lines superimposed on a linear background fitted across the
wavelength region (i.e., 1333.01$-$1355.55{\AA}) covered by the four
IRIS spectral windows. Table~\ref{table:tied} lists the lines used
in the fitting procedure. Lines with fixed positions are indicated
by a superscript `1'.  The two \SiII\ lines, indicated with a
superscript `2', are constrained to match the shift of the unblended
\SiII\ at 1350.06~\AA. The widths are fixed or constrained as given
in column 4. For example, the \SiII\ 1353.72 line has a maximum
width of 260 m{\AA}; while the line width of \FeII\ 1353.02 is fixed
at 41 m{\AA}.  The peak intensities of the blending lines are forced
to be in a fixed ratio (column 6) with the lines that they are tied
to (column 5). For example, the peak intensity of the blending line
\SiII\ 1353.72 is tied to the emission line \SiII\ 1350.06 in the
`\FeXII' window, and the intensity ratio is 0.49. The positions and
intensity ratios of the lines were determined by fitting Gaussians
to all the narrow (identified and unidentified) lines in 17 spectra
in the dataset with bright molecular and chromospheric lines but no
\FeXXI\ emission. For each line, the median position from the 17
spectral fits was taken as the line's fixed position. Then for each
narrow line blending with the \FeXXI\ we calculated its intensity
ratio with respect to all unblended narrow lines. We used the
variation in the intensity ratios to identify which line each
blended lines should be tied to, and the median of the ratio was
used to determine the strength of the blended line from the
intensity of its tied line. The lines used in the final fitting
procedure to obtain the \FeXXI\ intensities are marked and the
regions of background are indicated by the red-dashed lines in
Fig.~\ref{goodfits}. At most positions the fits appear to be
successful.

\begin{table}
\caption{The 17 emission lines from the four IRIS spectral windows
used in the multi-Gaussian fit.} \centering
\setlength{\tabcolsep}{1pt}
\begin{tabular}{c l c c c c}
 \hline\hline
IRIS window  &  Wavelength({\AA})  &  Ion  &  Width (m{\AA}) &  Intensity tied to  &  Ratio  \\
  \hline
            &   1333.45$^{1}$      & H$_2$   &     31      &       -                     &   -    \\
`\CII'      &   1333.79$^{1}$      & H$_2$, \SI\ &  41     &       -                     &   -     \\
  \hline
`1343'      &   1342.77$^{1}$      & H$_2$   &      31     &       -                     &   -     \\
  \hline
            &   1348.32$^{1}$      &   -     &      41     &       -                     &   -     \\
            &   1348.64$^{1}$      &   -     &      31     &       -                     &   -     \\
`\FeXII'    &   1349.63$^{1}$      &   -     &      41     &       -                     &   -     \\
            &   1350.06            & \SiII\  & $\leq$ 260  &       -                     &   -     \\
  \hline
            &   1352.74 $^2$       & \SiII\  & $\leq$ 260  &   \SiII\ 1350.06            &   0.54  \\
            &   1353.02$^{1}$      & \FeII\  &    41       &   \FeII\ 1354.76            &   1.85  \\
            &   1353.32$^{1}$      &    -    &    88       &   H$_2$  1342.77            &   0.79  \\
            &   1353.39$^{1}$      &    -    &    31       &   H$_2$  1342.77            &   1.50  \\
`\OI'       &   1353.72$^2$        & \SiII\  & $\leq$ 260  &   \SiII\ 1350.06            &   0.49  \\
            &   1354.01$^{1}$      & \FeII   &   41        &   \FeII\  1354.76           &   3.43  \\
            &   1354.08            & \FeXXI\ & $\geq$ 230  &      -                      &   -     \\
            &   1354.29            &  \CI\   & $\leq$ 130  &      -                      &   -     \\
            &   1354.76$^{1}$      & \FeII\  &   42        &      -                      &   -     \\
            &   1354.85$^{1}$      &  H$_2$  &   31        &   H$_2$  1342.77            &   2.0   \\
\hline \hline
\end{tabular}
\begin{tablenotes}
\item[1] 1: The emission lines with fixed positions in the
multi-Gaussian fit.
\item[2] 2: The emission lines with constrained positions in the multi-Gaussian
fit.
\end{tablenotes}
\label{table:tied}
\end{table}

\section{The Emission Measures of \FeXXI}
The observed intensity  is given by
\begin{eqnarray}
      I =  E_f\int{G(T,n_e)n_e^2dz}
\end{eqnarray}
where $n_e$ is the electron density, $T$ is electron temperature,
$z$ is the line-of-sight coordinate, $G(T,n_e)$ is the contribution
function of the line including the atomic abundance, and $E_f$ is
the instrument effective area or response function. The units of
$I$, in the equation above, depend on the units of the $E_f$. As
discussed below,  IRIS is a spectrometer and the intensities are
given in DN~s$^{-1}$. AIA is an imager and the intensity unit is
DN~px$^{-1}$~s$^{-1}$. To convert intensity to emission measure, we
need to divide by the respective response functions in the
appropriate units. Since the contribution functions of the IRIS and
AIA \FeXXI\ lines are sharply peaked in temperature and are formed
over the same heights, the contribution functions can be removed
from the integral. The emission measure in the \FeXXI\ plasma can
then be calculated from
\begin{eqnarray}
EM = \int n_e^2dz = I/(E_f G(T_{peak}))\qquad \textrm{cm$^{-5}$}.
\end{eqnarray}

\begin{figure} %%%%%%%%%%%%%%%%%%
\centering
\includegraphics[width=\linewidth,clip=]{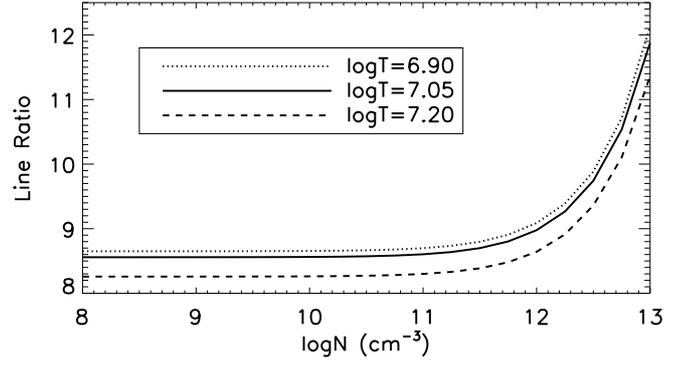}
\caption{The variation of the \FeXXI\ 128.75/1354.08 ratio as a
function of density, which is from CHIANTI V. 7.1.3
\citep{Landi13}.} \label{ratio}
\end{figure}

\begin{figure}    %%%%%%%%%%%%%%%%%%
\centering
\includegraphics[width=\linewidth,clip=]{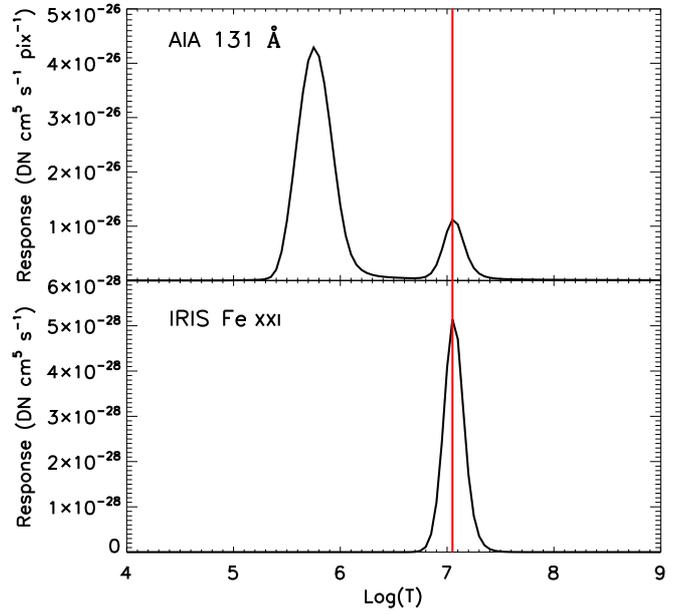}
\caption{Temperature response of  the AIA 131~\AA\ channel (top) and
the IRIS \FeXXI\ 1354~\AA\ line (bottom). The red lines are drawn
at the peak of the response curves, $11$~MK.} \label{rfs}
\end{figure}

\noindent The temperature response, $E_f G(T_{peak})$, for the AIA
131~\AA\ channel and the IRIS 1354~\AA\ line are shown in
Fig.~\ref{rfs}. The AIA function has been computed with the
procedure aia\_get\_response.pro with the calibration appropriate to
the time of the observation at the default pressure in CHANTI of
10$^{15}$~cm$^{-3}$~K \citep{Boerner12}. The keyword `chiantifix' is
also set to account for emission that is not included in the CHIANTI
database \citep{Testa12,DeRosa13}. The response of AIA to high and
low temperature flare emission is believed to be accurate to 25\%
with these empirical corrections \citep{Boerner12}. There are two
peaks: one due to \FeVIII\ formed around $6\times10^5$~K and the
other due to \FeXXI\ formed around $10^7$~K. The main variation in
the contribution function around 10~MK comes from the dependence of
ionization on temperature. The IRIS \FeXXI\ line is a fine structure
transition within the ground state so its excitation rate is
independent of temperature but at high densities ($n_e >
10^{12}$~cm$^{-3}$) it suffers from collisional de-excitation.
Fig.~\ref{ratio} shows the variation of the 128.75/1354.08 line
ratio, computed with CHIANTI \citep{Landi13}, as a function of
density for three temperatures close to the peak of the contribution
function. The temperature variation at low densities varies by less
than 7\% over the range shown. Only at temperatures below 4~MK is
there a significant temperature dependence on the ratio. We assumed
the low density limit and temperature 11~MK and computed the \FeXXI\
response function as the line emissivity divided by the radiometric
conversion coefficient, 2960 \citep{Depontieu14}, and the spectral
scale (0.0128~\AA). Again, the electron pressure was set to
10$^{15}$~cm$^{-3}$~K. The radiometric conversion for the IRIS
spectra is based on the International Ultraviolet Explorer (IUE)
spectral radiances, which has an uncertainty of 10$-$15\% (one
$\sigma$). Both the AIA and IRIS emissivities are computed at the
same pressure, elemental abundance and ionization. In the analysis
of the \FeXXI\ contribution to the 131 channel, we assume that the
ratio of the intensities is constant.

\FeXXI\ emission measure maps can be obtained by dividing the
observed intensities (DN~s$^{-1}$) by the value of the peak
response, at $11$~MK, of the respective functions. To compare the
IRIS and AIA results pixel-by-pixel pseudo rasters were constructed
from a series of AIA images by selecting, for each IRIS raster
position, the overlapping AIA pixels from the closest in time AIA
image. Along the slit, each AIA pixel (0.6\arcsec) is about 3.6 IRIS
pixels (0.167\arcsec). We map the IRIS pixels onto the AIA grid and
average the IRIS intensity in each AIA pixel. The IRIS raster step
size (1.01\arcsec) is larger than the AIA pixels so in the raster
direction the  single overlapping IRIS pixel is used. Thus, the IRIS
data have been  averaged across $\sim$3.6 pixels in the slit
direction, so that the IRIS data are re-binned to the scale of the
AIA data. In the raster direction, the AIA pseudo rasters are taken
from the closest time AIA data.

\section{Results}
\subsection{Images and spectra}
\begin{figure}    %%%%%%%%%%%%%%%%%%
\centering
\includegraphics[width=\linewidth,clip=]{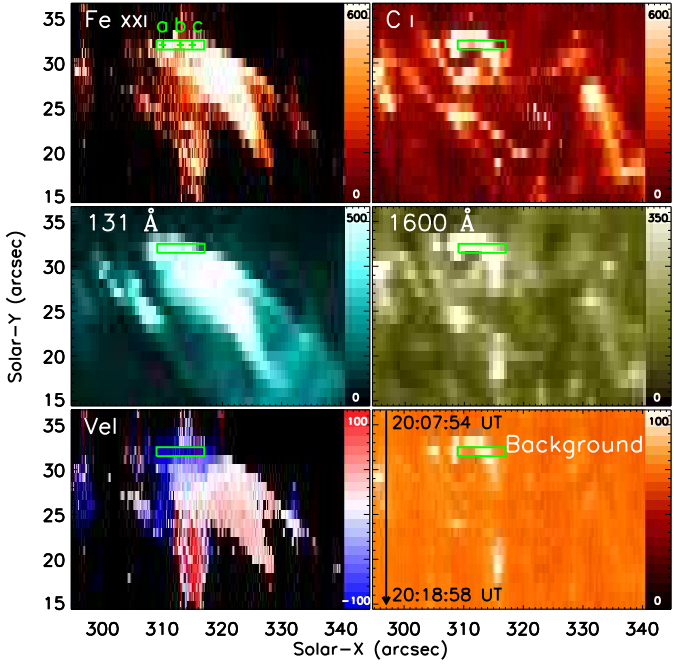}
\caption{Images of the first flare: (top) IRIS \FeXXI\ 1354.08~\AA\
and \CI\ 1354.29~\AA\ intensity; (middle) AIA 131~\AA\ and 1600~\AA\
pseudo-raster intensity; (bottom) IRIS \FeXXI\ Doppler velocity and
continuum intensity. The green rectangle marked with `+'s indicates
the region and sites of the spectra shown in Fig.~\ref{spectra1}.
The arrow indicates the time direction, and the duration of these
images is 11 m 4 s. The duration of the raster is marked in
Fig.~\ref{fig_goes}.} \label{pseudo1}
\end{figure}

\begin{figure}    %%%%%%%%%%%%%%%%%%
\centering
\includegraphics[width=\linewidth,clip=]{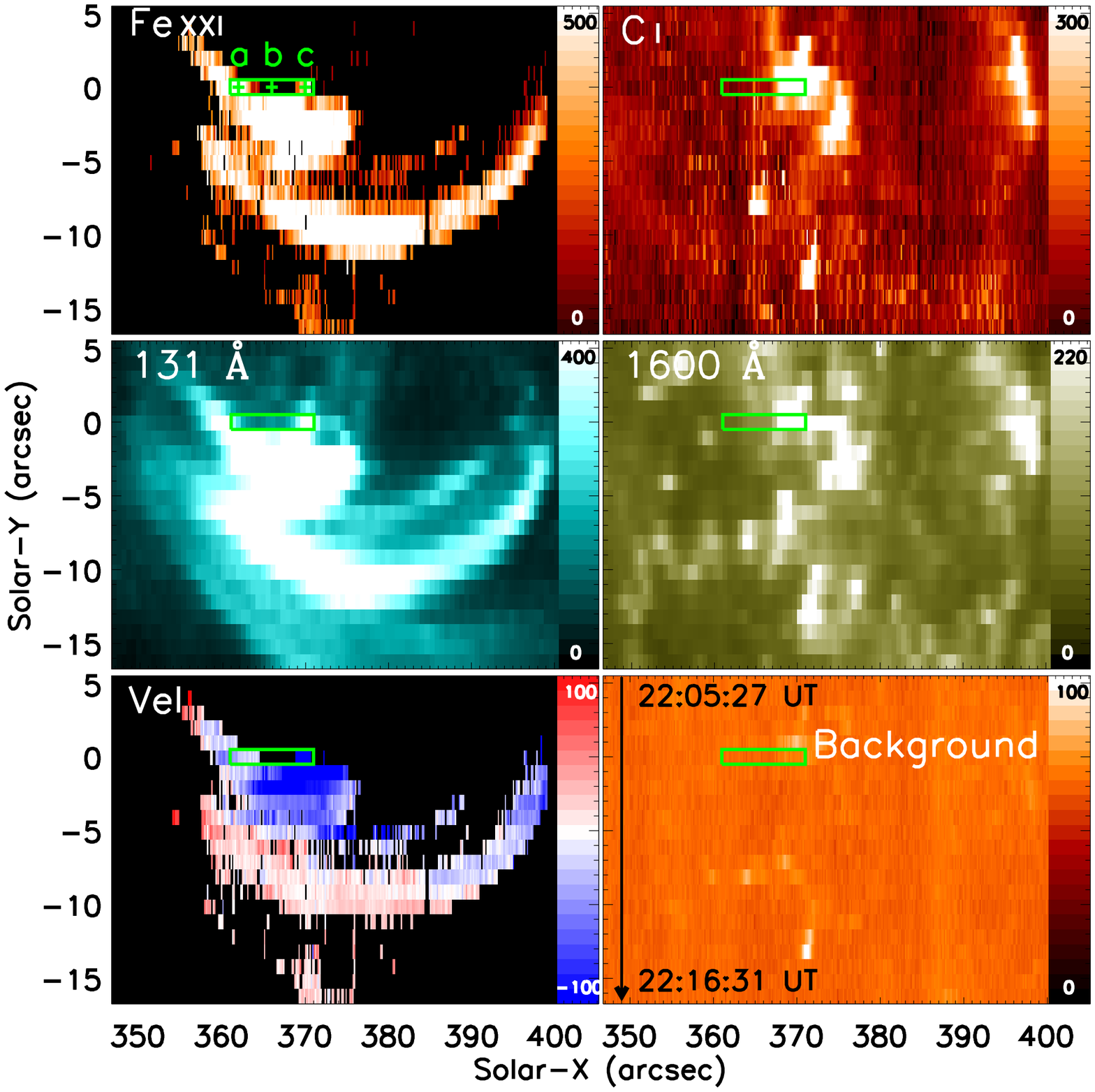}
\caption{Images of the second flare. (top) IRIS \FeXXI\ 1354.08~\AA\
and \CI\ 1354.29~\AA\ intensity; (middle) AIA 131~\AA\ and 1600~\AA\
pseudo-raster intensity; (bottom) IRIS \FeXXI\ Doppler velocity and
continuum intensity. The green rectangle marked with `+'s indicates
the region and sites of the spectra shown in Fig.~\ref{spectra2}.
The arrow indicates the time direction, and the duration of these
images is 11 m 4 s. The duration of the raster is marked in
Fig.~\ref{fig_goes}.} \label{pseudo2}
\end{figure}

There were two well-observed solar flares, as shown in
Fig.~\ref{fig_over}. The IRIS spectra of the two flares were fitted,
as described, with 17 Gaussians superimposed on a linear background.
The primary lines of interest are the \FeXXI\ 1354.08~\AA\ and, to a
lesser extent, \CI\ 1354.29~\AA. Figs.~\ref{pseudo1} and
\ref{pseudo2} compare IRIS \FeXXI\ and \CI\ intensity images with
AIA 131 and 1600~\AA\ pseudo-rasters of flare~1 and flare~2. The
regions of enhanced \CI\ identify the flare kernels. The \FeXXI\
Doppler velocity is shown in the bottom left panel. From the
velocity images we can see that the \FeXXI\ is blue shifted near the
start of the flare near the flare kernels. As shown, bright \CI\
overlaps very well with the bright 1600~\AA\ emission. The highest
velocity measured is about 200~\kms.

\begin{figure}    %%%%%%%%%%%%%%%%%%
\centering
\includegraphics[width=\linewidth,clip=]{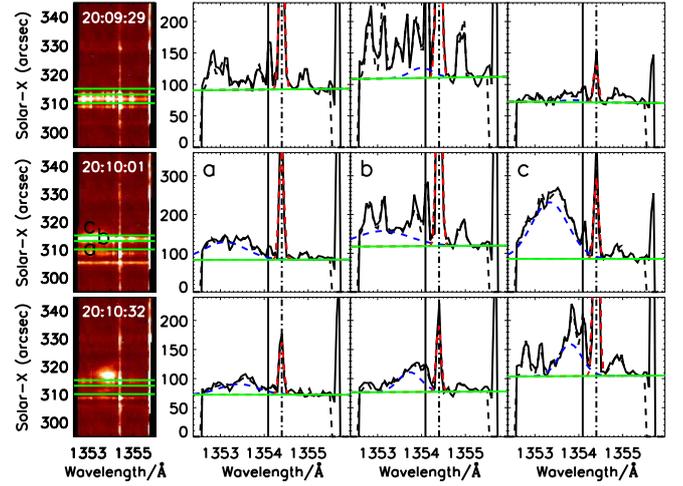}
\caption{Flare-kernel spectra from the region inside the green
rectangle on Fig.\ref{pseudo1}: (left-middle) the spectrogram of the
region marked with a green rectangle in Fig.~\ref{pseudo1}, with
(left-top) before and (left-bottom) after spectrograms of it. The
right panels show the spectra  (solid) and their fits (dashed) at
the positions marked by the green lines in the left panel. The
vertical solid lines mark the rest wavelength of \FeXXI\, and the
vertical dashed-dot lines mark the rest wavelength of \CI\ line. The
green solid line is the background, and the blue dashed line is the
background + \FeXXI\ line fit, the red dashed line is the background
+ \CI\ line fit.} \label{spectra1}
\end{figure}

\begin{figure}    %%%%%%%%%%%%%%%%%%
\centering
\includegraphics[width=\linewidth,clip=]{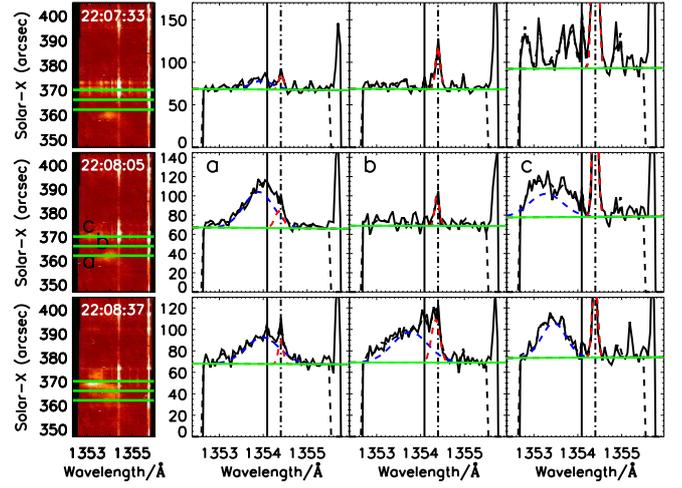}
\caption{Flare-kernel spectra from the region inside the green
rectangle on Fig.\ref{pseudo2}: (left-middle) the spectrogram at the
time of the rectangle region in Fig.~\ref{pseudo2}, with (top)
before and (bottom) after spectrogram. Right: spectra and their fits
(dashed) at the positions marked by the green lines in the left
panel. The vertical solid lines indicate the rest wavelength of
\FeXXI, and the vertical dashed-dot lines indicate the rest
wavelength of \CI. The green solid line is the background, and the
blue dashed line is the background + \FeXXI\ line fit, the red
dashed line is the background + \CI\ line fit.} \label{spectra2}
\end{figure}

To look closely at the emission from the kernels, we extract spectra
from almost neighboring positions along the slit. The points are
marked by a plus (`+') in Figs~\ref{pseudo1} and \ref{pseudo2}. In
Figs~\ref{spectra1} (flare~1) and \ref{spectra2} (flare~2), we show
spectral images and spectra from the first raster step with \FeXXI\
kernel emission, as well as spectra taken at the same positions from
the earlier and later raster steps. The resultant fits appear to be
very good. In both flares, the \FeXXI\ appears as a broad line below
the narrow chromospheric lines. The main sources of uncertainty in
the \FeXXI\ intensity come from the background fit, the edge of the
spectral window, and the possibly non-Gaussian \FeXXI\ profile. The
background is fitted across all spectral windows so is fairly well
constrained, and the window edge only affects the few positions
where \FeXXI\ is blue-shifted to by more than 150~\kms.

As well as assessing the fits by eyes we also computed the $\chi^2$
for the spectral region around the \FeXXI\ line. The values were
generally less than one. At one or two positions, the $\chi^2$ was
large ($> 10$). Here the peak of the \CI\ line could not be fit
correctly by a Gaussian possibly due to optical depth effects.
Occasionally individual narrow lines blending with \FeXXI\ caused
$\chi^2$  of about 3 at the position of the \FeXXI\ line. Since
these were much narrower than the \FeXXI\ they did not affect the
\FeXXI\ intensity.

\subsection{The EM of AIA 131~\AA\ and \FeXXI}

\begin{figure}    %%%%%%%%%%%%%%%%%%
\centering
\includegraphics[width=\linewidth,clip=]{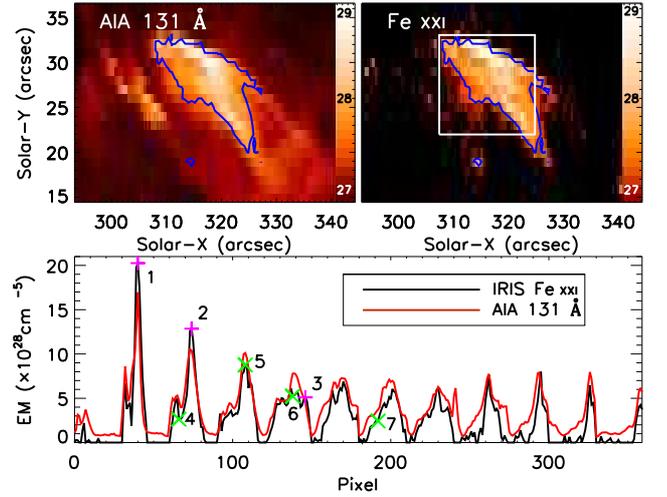}
\caption{Upper: The EMs computed from  AIA 131~\AA\ and IRIS \FeXXI\
at 11 MK for the first flare. The white box is the region over which
we compare EMs in the panels below. Bottom: Pixel-by-pixel and
row-by-row EMs across the white box region, starting from the upper
left. The black line is the \FeXXI\ EM, and the red line is the
131~\AA\ EM. The plus (`+') indicates those pixels where the \FeXXI\
emission is higher , and `$\times$' marks the selected pixels, shown
in Fig.~\ref{emsp1}} \label{em1}
\end{figure}

\begin{figure} %%%%%%%%%%%%%%%%%%
\centering
\includegraphics[width=\linewidth,clip=]{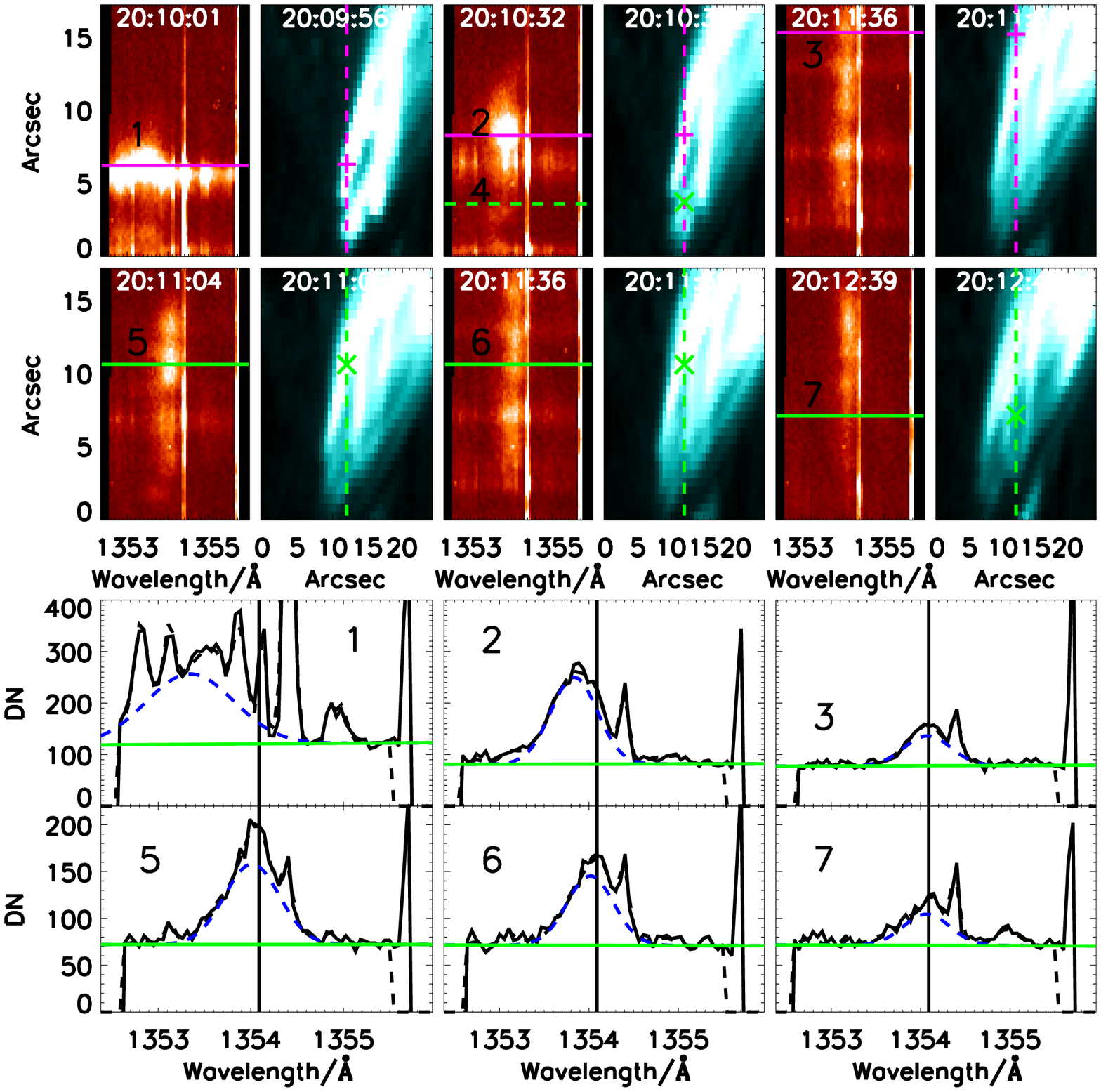}
\caption{Selected IRIS spectra and AIA intensities for the first
flare. The vertical dashed lines in the AIA images are the positions
of the IRIS slits. The numbers on the IRIS images show the positions
of the selected points in Fig.~\ref{em1}. The `+' and `$\times$' in
the AIA images mark the positions of these points in the AIA images.
Underneath are the IRIS spectra (solid black line) and their fits
(black dashed line) for the selected positions. The green solid line
is the background, the blue dashed line is background + \FeXXI\
line, while the vertical line is the rest wavelengths of the \FeXXI\
line.} \label{emsp1}
\end{figure}

\begin{figure}    %%%%%%%%%%%%%%%%%%
\centering
\includegraphics[width=\linewidth,clip=]{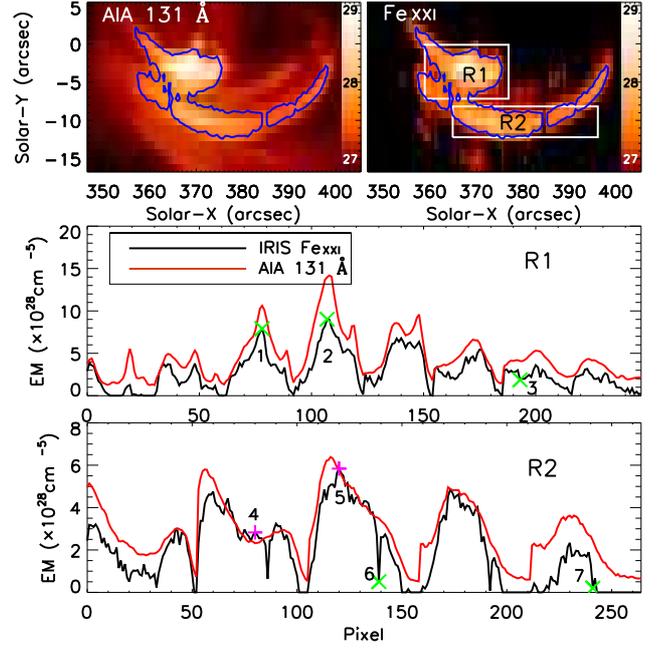}
\caption{The EMs from  AIA 131~\AA\ and IRIS \FeXXI\ at 11 MK for
the second flare. The two white boxes are the selected regions (R1
and R2), respectively. In the lower panels, the black line is the
\FeXXI\ EM, and the red line is the 131~\AA\ EM. The pluses (`+')
mark  pixels where the \FeXXI\ EM is higher , and the `$\times$'
marks selected pixels, shown in Fig.~\ref{emsp2} with higher 131
EM.} \label{em2}
\end{figure}

\begin{figure}
\centering
\includegraphics[width=\linewidth,clip=]{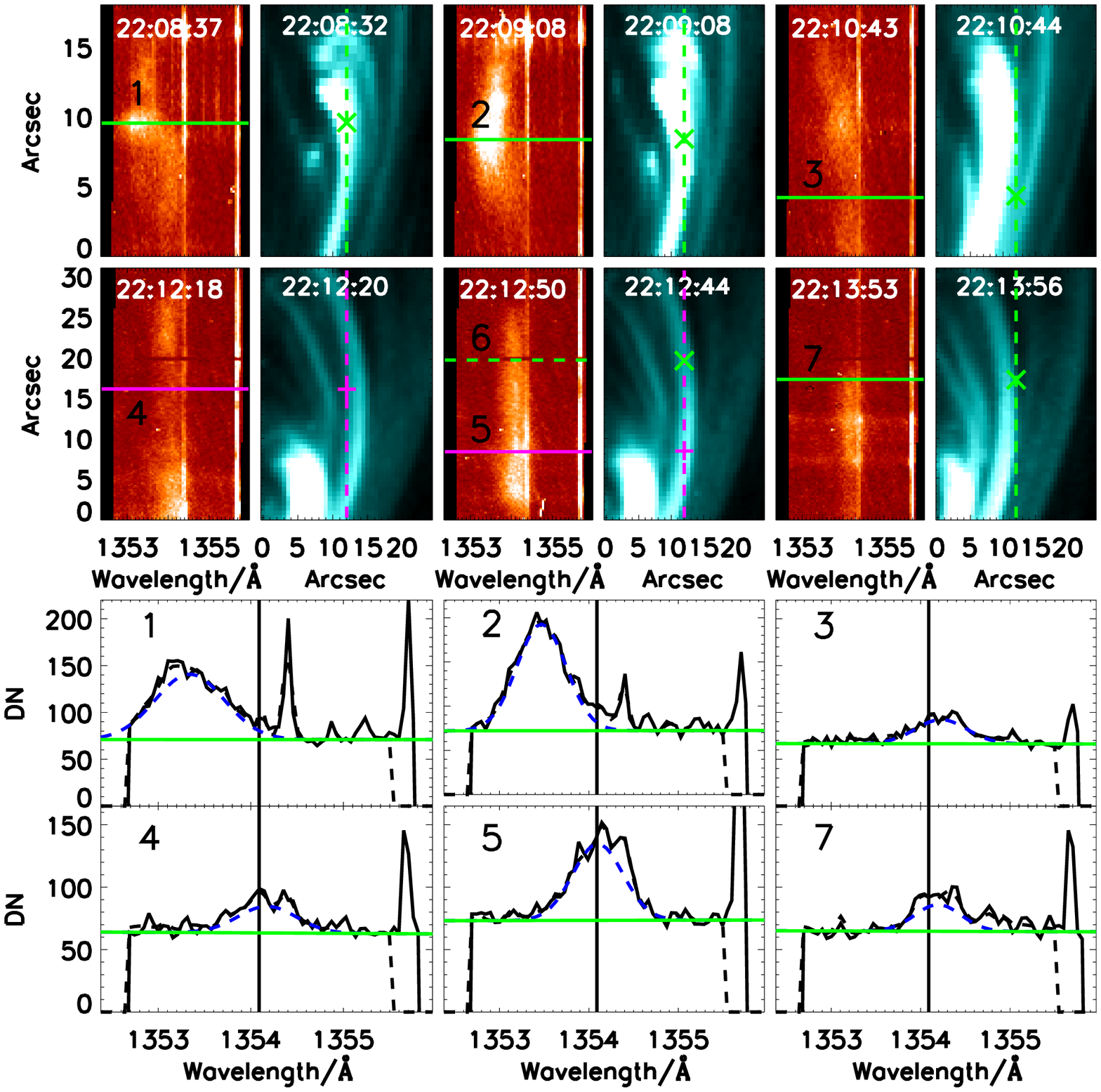}
\caption{Selected IRIS spectra and AIA intensities for the second
flare. The vertical lines in the AIA images mark the positions of
the IRIS slits. The numbers on the IRIS images show the positions of
the selected points in Fig.~\ref{em2}. The `+' and `$\times$' in the
AIA images mark the positions of these points in the AIA images.
Underneath are the IRIS spectra (solid black line) and their fits
(black dashed line). The green solid line is the background, the
blue dashed line is background + \FeXXI\ line, while the vertical
line is the rest wavelength of \FeXXI\ line. } \label{emsp2}
\end{figure}

From equation (2), we can estimate the EMs of the AIA 131 and IRIS
\FeXXI\ emitting plasma. Figs.~\ref{em1} shows the result for the
first flare. The upper panels are the images of the EMs with
contours of the IRIS \FeXXI\ EM. The images of the EMs have a
logarithmic scaling, and show good qualitative agreement. The lower
panel shows the comparison of the two EMs. To look closer at the
overlap, we  select the smaller region, surrounded by the white box,
and compare row-by-row the EMs of AIA 131 and \FeXXI\ in the panel
underneath. The structures are well matched. The AIA 131 EMs are
more diffuse and greater than \FeXXI\ EMs except around positions 1
and 2. As a check on the Gaussian line fit, we show the spectra and
fits at the numbered positions in Fig.~\ref{emsp1}. All fits appear
to be very good and there is no special feature of position 1 or 2
to indicate why the computed \FeXXI\ EM is more than the 131 EM.

We checked whether the higher \FeXXI\ than 131 EMs were due to
temporal mismatch between IRIS and AIA. AIA produced two 131~\AA\
exposures during each IRIS exposure (30 s) so we compared with both
of the overlapping AIA images separately for these and all other
points and found only very small differences. We therefore conclude
that it is not a temporal effect. It could  be an effect of the two
different spatial resolutions. The IRIS slit only samples about half
of the AIA pixel so if the features are very concentrated, then the
filling factor may be less in AIA than in IRIS. This would result in
a lower AIA than IRIS EM.

Similar plots for the second flare are shown in Figs~\ref{em2} and
\ref{emsp2}. The region of bright \FeXXI\ is slightly larger in the
second flare so we split the region in two, labeled `R1' and `R2',
in Fig.~\ref{em2}. Again selected spectra are shown (see,
Fig.~\ref{emsp2}). The overall behavior is similar to that seen in
the first flare: the 131 EMs are higher and more diffuse than the
\FeXXI\ EMs. At position 4, the IRIS EM is larger than the 131. As
shown in Fig.~\ref{emsp2}, this point is also on the edge of a loop,
so it may again be indicating a slight coalignment mismatch. We also
notice that at position 2 which is near the peak in the 131 emission
the IRIS EM is only about two thirds of the 131 EM. As seen in
Fig.~\ref{emsp2}, the  IRIS spectra at this position show a blue
shift of about 0.5~\AA\ or 100~\kms. Blue shifts in flares are
usually linked to chromospheric evaporation which implies that one
may expect a range of ionization states, including \FeVIII\ at this
position. Further support for this conclusion comes from the time
evolution that showed brightening at this site in all AIA channels
at the time of the IRIS observations.

\begin{figure} %%%%%%%%%%%%%%%%%%
\centering
\includegraphics[width=\linewidth,clip=]{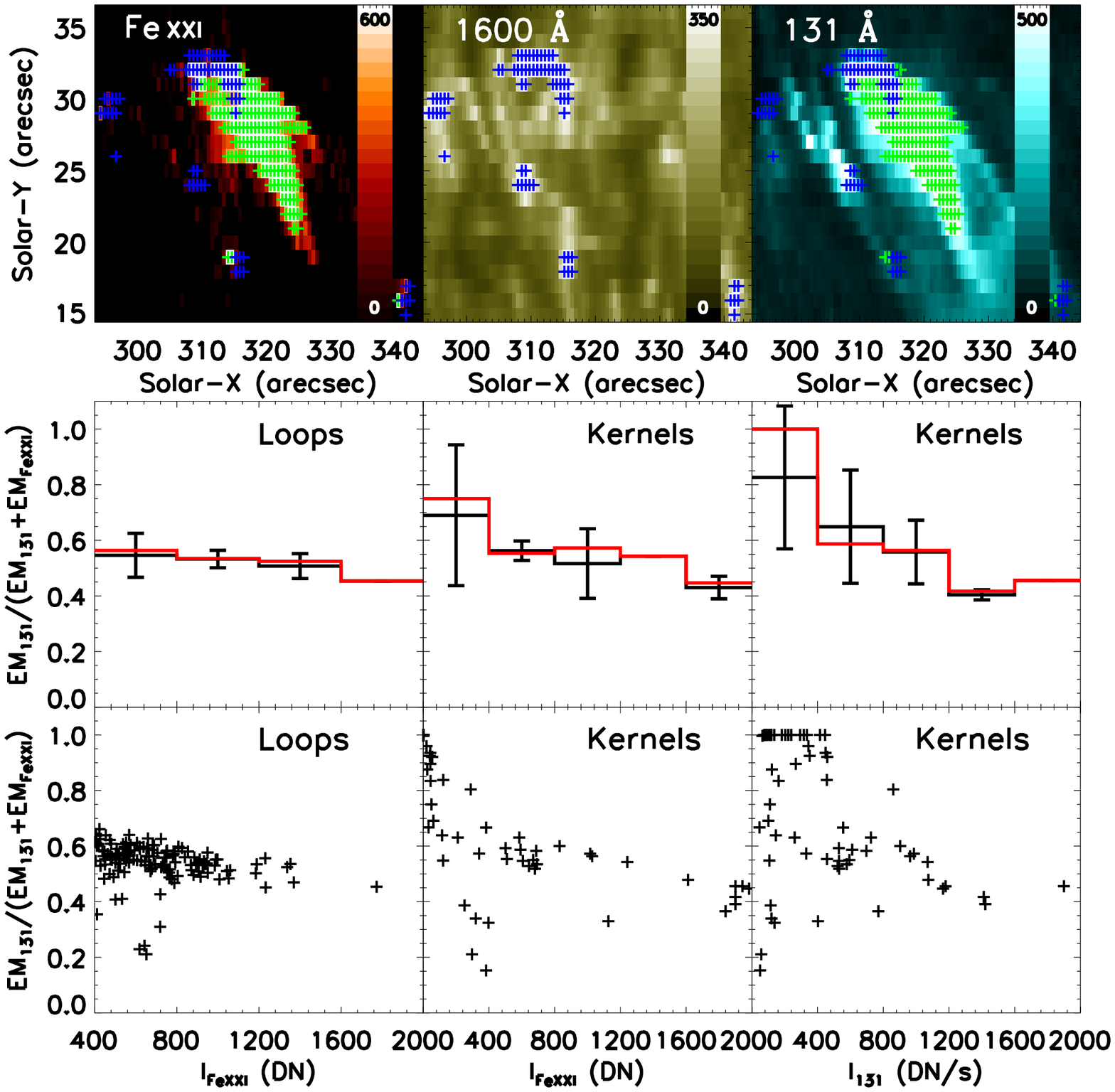}
\caption{Quantitative analysis of the \FeXXI\ and 131~\AA\ EMs for
flare1. The top row from left to right shows \FeXXI, 1600~\AA\ and
131~\AA\ intensity. The middle panels show the mean and median (red
line) ratios EM(131)/(EM(131)+EM(\FeXXI)) vs \FeXXI\ intensity in
loops (bin = 400~s$^{-1}$) and kernels (bin = 400~s$^{-1}$), and on
the right the ratio EM(131)/(EM(131)+EM(\FeXXI)) vs intensity of
131~\AA\ (bin = 400~s$^{-1}$). The bottom panels show the same
ratios vs intensity but as scatter plots. In the loops, the mean,
standard deviation and median ratios are 0.54$\pm$0.07, and 0.55.}
\label{fr6}
\end{figure}

\begin{figure} %%%%%%%%%%%%%%%%%%
\centering
\includegraphics[width=\linewidth,clip=]{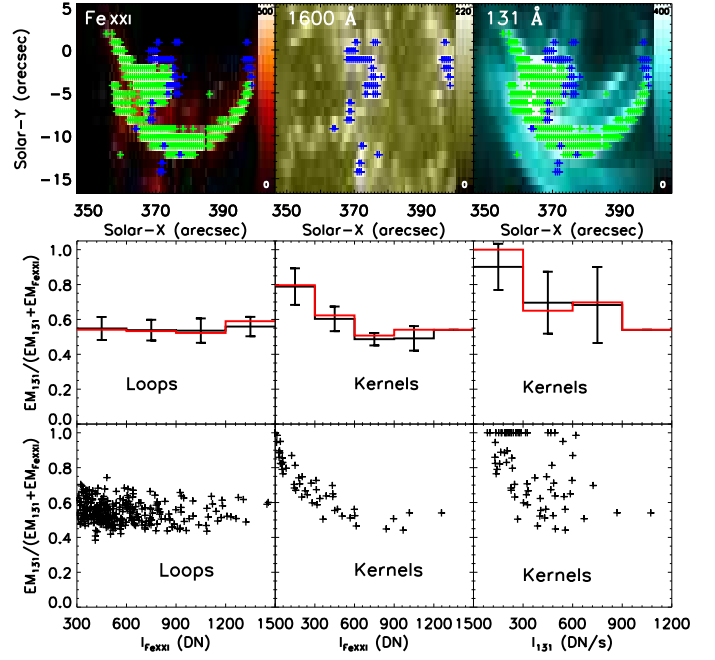}
\caption{Quantitative analysis of the \FeXXI\ and 131 EMs for flare
2. The top row from left to right shows \FeXXI, 1600~\AA\ and
131~\AA\ intensity. The middle panels show the mean and median (red
line) ratios EM(131)/(EM(131)+EM(\FeXXI)) vs \FeXXI\ intensity in
loops (bin = 300~DN~s$^{-1}$) and kernels (bin = 300~DN~s$^{-1}$),
and on the right the ratio EM(131)/(EM(131)+EM(\FeXXI)) vs intensity
of 131~\AA\ (bin= 300~DN~s$^{-1}$). The bottom panels show the same
ratios vs intensity but as scatter plots. In the loops, the mean,
standard deviation and median ratios are 0.55$\pm$0.06, and 0.54.}
\label{fr9}
\end{figure}

To obtain quantitative values we separate the two flare regions into
loops and kernels and compare the IRIS-to-AIA EM ratios as a
function of \FeXXI\ or 131 intensity. We plot the ratio of 131 EM to
the sum of 131 and \FeXXI\ EMs to avoid going off the scale when the
\FeXXI\ EM is very low. Thus when the 131 EM is due to \FeXXI\
alone, the ratio is 0.5. A contribution to the 131 EM from cooler
lines, continuum, or \FeXXIII\ would produce a value closer to one,
and where there is no \FeXXI\ contribution, the value is one. The
results for the two flares are shown in Figs.~\ref{fr6} and
\ref{fr9}. The selection of loop and kernel regions is based on the
AIA 131 and 1600~\AA\ intensities. Pixels with 131~\AA\ intensities
greater than twice 131~\AA\ average and 1600~\AA\ intensities less
than twice 1600~\AA\ average are selected as loops and those with
1600~\AA\ intensities greater than twice 1600~\AA\ average are
selected as kernels.

For loops the average value of the ratio is 0.54 in flare 1 and 0.55
in flare 2. The ratio is almost constant across the intensity range.
This implies that typically about 80-85\% of the 131~\AA\ loop
emission is from \FeXXI\ which is consistent with \citet{Milligan13}
who estimate a 20\% continuum contribution to the 131~\AA\ channel
during flares.

For flare 1, the ratio ranges from 0.2 to 0.6. The low values can be
attributed to alignment or filling factor issues. For example the
single point in the highest intensity bin in flare 1 is from a
bright point on the edge of the kernels. The two points in the next
highest intensity bin with a ratio below 0.5 are also from the edge
of the kernel region. The other points with ratios less than 0.5 are
from the bright western edge of the 131~\AA\ region where both the
\FeXXI\ and 131~\AA\ emission rapidly decreases. The higher ratio
values are mostly from the lowest intensity bin on the edge of the
loop regions. These may be due to cooler material on the edge of the
loop or background 131~\AA\ emission. Therefore the typical loop
value in flare 1 is the mean of the low intensity bins, 0.55 or 81\%
\FeXXI.

The flare 2 loops have more scatter in their ratio. In the highest
intensity bin, there are four points with a ratio around 0.6 (67\%
\FeXXI). These are from the bright region at the center of R1. As
mentioned, these points have blueshifted \FeXXI\ and at the same
time brightened in all AIA channels, suggesting it is a site of
chromospheric evaporation, not a typical loop. The scatter in the
low intensity bins is mostly caused by coalignment issues in the R2
region as can be seen in Fig.~\ref{em2}. The last row, around
position 7, produced many of the high ratio points. These points are
along the edge of the loop and may, similar to the edges of the
loops in flare 1, maybe due to cooler material on the edge of the
loop or background 131~\AA\ emission. In flare 2, the average value
0.55 (81\% \FeXXI\ is typical for the 131~\AA\ loops.

The ratio in the kernels covers the full range from $0-100$\%
\FeXXI. We have plotted the ratio as a function of both \FeXXI\ and
131 intensity. As the intensities increase the contribution of
\FeXXI\ to the 131~\AA\ channel increases up to the value seen in
the loops. In flare 1, as noted above and shown in Fig.~\ref{fr6},
the \FeXXI\ EM is greater than the 131 EM in the brightest pixels.
We attribute this to imperfect coalignment or different filling
factors for the two instruments, so the ratio in the two highest
intensity bins of flare 1 are not included in our final conclusions.
Below a 131~\AA\ intensity of 600~DN~s$^{-1}$, a significant
fraction of  pixels have no \FeXXI\ emission. Thus there is
structure within the kernels on the scale of the AIA pixel size,
0.6\arcsec. Considering the 131 pixels in the middle of the
intensity range, one sees from Figs.~\ref{fr6} and \ref{fr9} that
the ratio is between 0.55 and 0.75 which implies the 131 kernel EM
is between 20-60\% greater than the \FeXXI\ EM. This broad range
again implies high resolution structure in the kernels not seen in
the 131~\AA\ images. If the continuum emission is about 20\% of the
\FeXXI\ as found by \citet{Milligan13}, then our result implies from
zero to 52\% of the flare kernel 131~\AA\ emission is due to cooler
plasma emission.

These results assume that both instruments are well calibrated and
the data are coaligned. The fact that we obtain a consistent value
of 80\% \FeXXI\ contribution to the 131 channel in the loops
suggests that the calibration is good. A 20\% error in the
intercalibration would lead to a 10\% change in the \FeXXI\
contribution.

\section{Conclusions and Discussions}
Using the high resolution IRIS spectral data, together with the SDO
data, we study two solar flares in AR 11875 that occurred on 24
October 2013. We obtain the intensity of \FeXXI\ from IRIS data and
compare it with AIA data at 131~\AA. Coalignment to an accuracy of
one AIA pixel (0.6\arcsec) is achieved by coaligning the IRIS \CI\
to AIA 1600~\AA\ (see Fig.~\ref{fig_coalign}) and the IRIS \FeXXI\
to AIA 131~\AA\ images. \FeXXI\ is a hot and broad line which, in
flare kernels, is blended with many cold lines. Previous studies
\citep{Doschek75,Cheng79,Mason86} at lower resolution only consider
blending of \FeXXI\ with the \CI\ line. However, as IRIS reveals,
there are several additional cold chromospheric lines blending with
the \FeXXI\ (see Fig.~\ref{goodfits}). After studying other IRIS
flare kernel spectra, we were able to identify lines in different
parts of the IRIS spectrum that can be used to constrain the
intensity, widths and positions of the blending cold lines (see
Table~1). Using 17 Gaussian lines and a linear continuum we are able
to fit the spectra and obtain reliable \FeXXI\ intensities at  all
positions.

The spectra show regions at the onset of the flares where of \FeXXI\
is coincident with cold-line emission (see Fig.~\ref{pseudo1}
and~\ref{pseudo2}). We have carefully checked the fits to make sure
that the procedure has worked at these positions (see
Fig.~\ref{spectra1} and~\ref{spectra2}). In these regions, the
\FeXXI\ line shows  high blue shifts up to 200~\kms, which is the
edge of the spectral window.  This velocity is also consistent with
other hot spectral lines seen in flares, such as Fe XIX
\citep{Czaykowska01,Teriaca03,Brosius04}, Fe XXIII and Fe XXIV
\citep{Milligan09}. In our study, the EMs have the magnitude of
about 10$^{28}$ cm$^{-5}$, some pixels can reach 10$^{29}$ cm$^{-5}$
(see Fig.~\ref{em1} and~\ref{em2}), this also agrees with others
\citep{Graham13,Fletcher13}.

It is well known that the 131~\AA\ channel in AIA  has contributions
from the \FeXXI\ and the \FeVIII\ line. \citet{Brosius12} list a
number of other transition region lines that may also be present. We
compare the EMs of AIA 131~\AA\ and the IRIS \FeXXI\ line
pixel-by-pixel, assuming all emission is from hot plasma.  For loop
regions the AIA 131~\AA\ EMs are about 20\% greater than \FeXXI\ EMs
(see Fig.~\ref{fr6} and~\ref{fr9}). This is consistent with the
analysis of \citet{Milligan13} that 20\%\ of the 131~\AA\ channel
emission is due to continuum in flares. We also identified a loop
region with about 33\% greater 131 than \FeXXI\ EM. Since, this
region simultaneously brightened in all AIA channels and was
associated with \FeXXI\ blue shifts, we attribute the lower than
average \FeXXI\ to additional \FeVIII\ due to a recent injection of
cool plasma into the loop. In the brighter 131 flare kernels ($>
600$~DN~s$^{-1}$), the 131 EM is about 20-60\% more than the IRIS
\FeXXI\ suggesting that there is an up to 52\% contribution from
cool plasma emission \citep{Brosius12}.

Small-scale kernel structure results in a broad range of 131/\FeXXI\
EM ratios and sharp gradients in IRIS \FeXXI\ emission at sites of
molecular and transition region emission. At the start of the flare,
consecutive raster steps showed a large increase in \FeXXI\ across
the kernels indicating that the kernel structure is both temporal
and spatial. Future work will focus on the analysis of flares
obtained at higher spectral resolution with broader wavelength
windows and higher temporal cadence to resolve the temporal/spatial
ambiguities.

\begin{acknowledgements}
The authors would like to thank the anonymous referee for valuable
comments to improve the manuscript. IRIS is a NASA small explorer
mission developed and operated by LMSAL with mission operations
executed at NASA Ames Research center and major contributions to
down-link communications funded by the Norwegian Space Center (NSC,
Norway) through an ESA PRODEX contract. We are indebted to the IRIS
and SDO teams for providing the high resolution data. We would also
thank the people at MPS, in particular to Professor Hardi Peter,
Doctors Li, Leping, Chen, Feng and Guo, Lijia.
\end{acknowledgements}


\begin{thebibliography} {38}

\bibitem[{{Antonucci} {et~al.}(1982){Antonucci}, {Gabriel}, {Acton},
{Leibacher}, {Culhane}, {Rapley}, {Doyle}, {Machado}, \&
{Orwig}}]{Antonucci82} {Antonucci}, E., {Gabriel}, A.~H., {Acton},
L.~W., {et~al.} 1982, \solphys, 78, 107

\bibitem[{{Benz}(2008)}]{Benz08} {Benz}, A.~O. 2008, Living Reviews in Solar Physics, 5, 1

\bibitem[{{Boerner} {et~al.}(2012){Boerner}, {Edwards}, {Lemen}, {Rausch},
{Schrijver}, {Shine}, {Shing}, {Stern}, {Tarbell}, {Title},
{Wolfson}, {Soufli}, {Spiller}, {Gullikson}, {McKenzie}, {Windt},
{Golub}, {Podgorski}, {Testa}, \& {Weber}}]{Boerner12} {Boerner},
P., {Edwards}, C., {Lemen}, J., {et~al.} 2012, \solphys, 275, 41

\bibitem[{{Brosius} \& {Holman}(2012)}]{Brosius12}
{Brosius}, J.~W. \& {Holman}, G.~D. 2012, \aap, 540, A24

\bibitem[{{Brosius} \& {Phillips}(2004)}]{Brosius04}
{Brosius}, J.~W. \& {Phillips}, K.~J.~H. 2004, \apj, 613, 580

\bibitem[{{Cheng} {et~al.}(1979){Cheng}, {Feldman}, \& {Doschek}}]{Cheng79}
{Cheng}, C.-C., {Feldman}, U., \& {Doschek}, G.~A. 1979, \apj, 233,
736

\bibitem[{{Cheng} {et~al.}(2015){Cheng}, {Ding}, \& {Fang}}]{Cheng15}
{Cheng}, X., {Ding}, M.~D., \& {Fang}, C. 2015, \apj, 804, 82

\bibitem[{{Czaykowska} {et~al.}(2001){Czaykowska}, {Alexander}, \& {De
Pontieu}}]{Czaykowska01} {Czaykowska}, A., {Alexander}, D., \& {De
Pontieu}, B. 2001, \apj, 552, 849

\bibitem[{{De Pontieu} {et~al.}(2014){De Pontieu}, {Title}, {Lemen}, {Kushner},
{Akin}, {Allard}, {Berger}, {Boerner}, {Cheung}, {Chou}, {Drake},
{Duncan}, {Freeland}, {Heyman}, {Hoffman}, {Hurlburt}, {Lindgren},
{Mathur}, {Rehse}, {Sabolish}, {Seguin}, {Schrijver}, {Tarbell},
{W{\"u}lser}, {Wolfson}, {Yanari}, {Mudge}, {Nguyen-Phuc},
{Timmons}, {van Bezooijen}, {Weingrod}, {Brookner}, {Butcher},
{Dougherty}, {Eder}, {Knagenhjelm}, {Larsen}, {Mansir}, {Phan},
{Boyle}, {Cheimets}, {DeLuca}, {Golub}, {Gates}, {Hertz},
{McKillop}, {Park}, {Perry}, {Podgorski}, {Reeves}, {Saar}, {Testa},
{Tian}, {Weber}, {Dunn}, {Eccles}, {Jaeggli}, {Kankelborg},
{Mashburn}, {Pust}, {Springer}, {Carvalho}, {Kleint}, {Marmie},
{Mazmanian}, {Pereira}, {Sawyer}, {Strong}, {Worden}, {Carlsson},
{Hansteen}, {Leenaarts}, {Wiesmann}, {Aloise}, {Chu}, {Bush},
{Scherrer}, {Brekke}, {Martinez-Sykora}, {Lites}, {McIntosh},
{Uitenbroek}, {Okamoto}, {Gummin}, {Auker}, {Jerram}, {Pool}, \&
{Waltham}}]{Depontieu14} {De Pontieu}, B., {Title}, A.~M., {Lemen},
J.~R., {et~al.} 2014, \solphys, 289, 2733

\bibitem[{{DeRosa} \& {Slater}(2013)}]{DeRosa13}
{DeRosa}, M. \& {Slater}, G. 2013, {Guide to SDO Data Analysis
edited on February 19, 2013}

\bibitem[{{Doschek} {et~al.}(1975){Doschek}, {Dere}, {Sandlin}, {Vanhoosier},
{Brueckner}, {Purcell}, {Tousey}, \& {Feldman}}]{Doschek75}
{Doschek}, G.~A., {Dere}, K.~P., {Sandlin}, G.~D., {et~al.} 1975,
\apjl, 196, L83

\bibitem[{{Feldman} {et~al.}(2000){Feldman}, {Curdt}, {Landi}, \&
{Wilhelm}}]{Feldman00} {Feldman}, U., {Curdt}, W., {Landi}, E., \&
{Wilhelm}, K. 2000, \apj, 544, 508

\bibitem[{{Fletcher} {et~al.}(2011){Fletcher}, {Dennis}, {Hudson}, {Krucker},
{Phillips}, {Veronig}, {Battaglia}, {Bone}, {Caspi}, {Chen},
{Gallagher}, {Grigis}, {Ji}, {Liu}, {Milligan}, \&
{Temmer}}]{Fletcher11} {Fletcher}, L., {Dennis}, B.~R., {Hudson},
H.~S., {et~al.} 2011, \ssr, 159, 19

\bibitem[{{Fletcher} {et~al.}(2013){Fletcher}, {Hannah}, {Hudson}, \&
{Innes}}]{Fletcher13} {Fletcher}, L., {Hannah}, I.~G., {Hudson},
H.~S., \& {Innes}, D.~E. 2013, \apj, 771, 104

\bibitem[{{Fletcher} \& {Hudson}(2001)}]{Fletcher01}
{Fletcher}, L. \& {Hudson}, H. 2001, \solphys, 204, 69

\bibitem[{{Fletcher} {et~al.}(2004){Fletcher}, {Pollock}, \&
{Potts}}]{Fletcher04} {Fletcher}, L., {Pollock}, J.~A., \& {Potts},
H.~E. 2004, \solphys, 222, 279

\bibitem[{{Gan} {et~al.}(2008){Gan}, {Li}, \& {Miroshnichenko}}]{Gan08}
{Gan}, W.~Q., {Li}, Y.~P., \& {Miroshnichenko}, L.~I. 2008, Advances
in Space Research, 41, 908

\bibitem[{{Graham} {et~al.}(2013){Graham}, {Hannah}, {Fletcher}, \&
{Milligan}}]{Graham13} {Graham}, D.~R., {Hannah}, I.~G., {Fletcher},
L., \& {Milligan}, R.~O. 2013, \apj, 767, 83

\bibitem[Graham \& Cauzzi(2015)]{Graham15} Graham, D.~R., \& Cauzzi, G.\ 2015,
\apjl, 807, L22

\bibitem[{{Innes}(2001)}]{Innes01}
{Innes}, D.~E. 2001, \aap, 378, 1067

\bibitem[{{Innes}(2008)}]{Innes08}
{Innes}, D.~E. 2008, \aap, 481, L41

\bibitem[{{Kliem} {et~al.}(2002){Kliem}, {Dammasch}, {Curdt}, \&
{Wilhelm}}]{Kliem02} {Kliem}, B., {Dammasch}, I.~E., {Curdt}, W., \&
{Wilhelm}, K. 2002, \apjl, 568, L61

\bibitem[{{Landi} {et~al.}(2013){Landi}, {Young}, {Dere}, {Del Zanna}, \&
{Mason}}]{Landi13} {Landi}, E., {Young}, P.~R., {Dere}, K.~P., {Del
Zanna}, G., \& {Mason}, H.~E. 2013, \apj, 763, 86

\bibitem[{{Lemen} {et~al.}(2012){Lemen}, {Title}, {Akin}, {Boerner}, {Chou},
{Drake}, {Duncan}, {Edwards}, {Friedlaender}, {Heyman}, {Hurlburt},
{Katz}, {Kushner}, {Levay}, {Lindgren}, {Mathur}, {McFeaters},
{Mitchell}, {Rehse}, {Schrijver}, {Springer}, {Stern}, {Tarbell},
{Wuelser}, {Wolfson}, {Yanari}, {Bookbinder}, {Cheimets},
{Caldwell}, {Deluca}, {Gates}, {Golub}, {Park}, {Podgorski}, {Bush},
{Scherrer}, {Gummin}, {Smith}, {Auker}, {Jerram}, {Pool}, {Soufli},
{Windt}, {Beardsley}, {Clapp}, {Lang}, \& {Waltham}}]{Lemen12}
{Lemen}, J.~R., {Title}, A.~M., {Akin}, D.~J., {et~al.} 2012,
\solphys, 275, 17

\bibitem[Li et al.(2015)]{Li15} Li, D., Ning, Z.~J.,
\& Zhang, Q.~M.\ 2015, \apj, 807, 72

\bibitem[{{Mason} {et~al.}(1986){Mason}, {Shine}, {Gurman}, \&
{Harrison}}]{Mason86} {Mason}, H.~E., {Shine}, R.~A., {Gurman},
J.~B., \& {Harrison}, R.~A. 1986, \apj, 309, 435

\bibitem[{{Mclntosh} {et~al.}(2013){Mclntosh}, {De Pontieu}, {Hansteen}, \&
{Boerner}}]{McIntosh13} {Mclntosh}, S.~W., {De Pontieu}, B.,
{Hansteen}, V., \& {Boerner}, P. 2013, {A User's Guide To IRIS Data
Retrieval, Reduction and Analysis, 2013, October, 30}

\bibitem[{{Milligan} \& {Dennis}(2009)}]{Milligan09}
{Milligan}, R.~O. \& {Dennis}, B.~R. 2009, \apj, 699, 968

\bibitem[{{Milligan} {et~al.}(2006){Milligan}, {Gallagher}, {Mathioudakis},
{Bloomfield}, {Keenan}, \& {Schwartz}}]{Milligan06} {Milligan},
R.~O., {Gallagher}, P.~T., {Mathioudakis}, M., {et~al.} 2006, \apjl,
638, L117

\bibitem[{{Milligan} \& {McElroy}(2013)}]{Milligan13}
{Milligan}, R.~O. \& {McElroy}, S.~A. 2013, \apj, 777, 12

\bibitem[{{Ning} \& {Cao}(2011)}]{Ning11}
{Ning}, Z. \& {Cao}, W. 2011, \solphys, 269, 283

\bibitem[O'Dwyer et al.(2010)]{ODwyer10} O'Dwyer, B., Del Zanna, G., Mason, H.~E.,
Weber, M.~A., \& Tripathi, D.\ 2010, \aap, 521, A21

\bibitem[{{Qiu} {et~al.}(2002){Qiu}, {Lee}, {Gary}, \& {Wang}}]{Qiu02}
{Qiu}, J., {Lee}, J., {Gary}, D.~E., \& {Wang}, H. 2002, \apj, 565,
1335

\bibitem[{{Teriaca} {et~al.}(2003){Teriaca}, {Falchi}, {Cauzzi}, {Falciani},
{Smaldone}, \& {Andretta}}]{Teriaca03} {Teriaca}, L., {Falchi}, A.,
{Cauzzi}, G., {et~al.} 2003, \apj, 588, 596

\bibitem[{{Teriaca} {et~al.}(2006){Teriaca}, {Falchi}, {Falciani}, {Cauzzi}, \&
{Maltagliati}}]{Teriaca06} {Teriaca}, L., {Falchi}, A., {Falciani},
R., {Cauzzi}, G., \& {Maltagliati}, L. 2006, \aap, 455, 1123

\bibitem[{{Testa} {et~al.}(2012){Testa}, {Drake}, \& {Landi}}]{Testa12}
{Testa}, P., {Drake}, J.~J., \& {Landi}, E. 2012, \apj, 745, 111

\bibitem[{{Tian} {et~al.}(2014){Tian}, {DeLuca}, \& {Reeves}}]{Tian14}
{Tian}, H., {DeLuca}, E., \& {Reeves}, K.~K., et al. 2014, \apj,
786, 137

\bibitem[Tian et al.(2015)]{Tian15} Tian, H., Young, P.~R.,
Reeves, K.~K., et al.\ 2015, \apj, 811, 139

\bibitem[{{Wang} {et~al.}(2003){Wang}, {Solanki}, {Curdt}, {Innes}, {Dammasch},
\& {Kliem}}]{Wang03} {Wang}, T.~J., {Solanki}, S.~K., {Curdt}, W.,
{et~al.} 2003, \aap, 406, 1105

\bibitem[{{Young} {et~al.}(2013){Young}, {Doschek}, {Warren}, \&
{Hara}}]{Young13} {Young}, P.~R., {Doschek}, G.~A., {Warren}, H.~P.,
\& {Hara}, H. 2013, \apj, 766, 127

\bibitem[Young et al.(2015)]{Young15} Young, P.~R., Tian, H.,
\& Jaeggli, S.\ 2015, \apj, 799, 218

\end{thebibliography}
\end{document}